\documentclass[10pt,aps,prd,superscriptaddress,nofootinbib,nobibnotes,longbibliography,floatfix,twocolumn]{revtex4-2}

\usepackage{bm}
\usepackage{mathtools,
amsmath,
amssymb,
amsfonts,
mathrsfs,
chngcntr,
multirow}

\let\cc\corresponds
\let\corresponds\relax
\usepackage{mathabx}
\let\corresponds\cc

\usepackage[utf8]{inputenc}
\usepackage[T1]{fontenc}

\usepackage{soul}

\usepackage[dvipsnames]{xcolor}
\usepackage[unicode]{hyperref}
\hypersetup{colorlinks=true, citecolor=MidnightBlue,
            linkcolor=MidnightBlue, urlcolor=MidnightBlue, linktocpage=true}
\usepackage[normalem]{ulem}

\usepackage{graphicx}

\begin{document}
\title{The dynamics of spherically symmetric black holes in scalar-Gauss-Bonnet gravity with a Ricci coupling}

\author{Farid~Thaalba}
\address{Nottingham Centre of Gravity,
Nottingham NG7 2RD, United Kingdom}
\address{School of Mathematical Sciences, University of Nottingham,
University Park, Nottingham NG7 2RD, United Kingdom}

\author{Nicola~Franchini}
\affiliation{Université Paris Cit\'e, CNRS, Astroparticule et Cosmologie,  F-75013 Paris, France}
\affiliation{CNRS-UCB International Research Laboratory, Centre Pierre Binétruy, IRL2007, CPB-IN2P3, Berkeley, CA 94720, USA}

\author{Miguel~Bezares}
\address{Nottingham Centre of Gravity,
Nottingham NG7 2RD, United Kingdom}
\address{School of Mathematical Sciences, University of Nottingham,
University Park, Nottingham NG7 2RD, United Kingdom}

\author{Thomas~P.~Sotiriou}
\address{Nottingham Centre of Gravity,
Nottingham NG7 2RD, United Kingdom}
\address{School of Mathematical Sciences, University of Nottingham,
University Park, Nottingham NG7 2RD, United Kingdom}
\address{School of Physics and Astronomy, University of Nottingham,
University Park, Nottingham NG7 2RD, United Kingdom}

\begin{abstract}
  We study the dynamics of spherically symmetric black holes in scalar Gauss-Bonnet gravity with an additional coupling between the scalar field and the Ricci scalar using non-linear simulations that employ excision. In this class of theories, black holes possess hair if they lie in a specific mass range, in which case they exhibit a finite-area singularity, unlike general relativity. 
    Our results show that the Ricci coupling can mitigate the loss of hyperbolicity in spherical evolution with black hole initial data.  Using excision can enlarge the parameter space for which the system remains well-posed, as one can excise the elliptic region that forms inside the horizon. Furthermore, we explore a possible relation between the loss of hyperbolicity and the formation of the finite-area singularity inside the horizon. We find that the location of the singularity extracted from the static analysis matches the location of the sonic line well. Finally, when possible, we extract the monopolar quasi-normal modes and the time scale of the linear tachyonic instability associated with scalarization. We also check our results by utilizing a continued fraction analysis and supposing linear perturbations of the static solutions.
\end{abstract}

\maketitle

\section{Introduction}
Compact objects such as neutron stars (NSs), black holes (BHs), and coalescing binaries thereof constitute an avenue to examine the gravitational interaction beyond our current understanding.
In general relativity (GR), electro-vacuum black holes are covered by various uniqueness theorems~\cite{Hawking:1971vc,Carter:1971zc,Israel:1967wq,Israel:1967za,Wald:1971iw,Mazur:1982db}. Moreover, scalar and vector fields minimally coupled to GR cannot dress a BH with extra charges~\cite{Chase:1970omy,Teitelboim:1972qx,Bekenstein:1972ny}. Additionally, BHs in scalar-tensor theories with self-interacting scalars are also protected by no-hair theorems~\cite{Sotiriou:2015pka,Herdeiro:2015waa,Sotiriou:2011dz,Bekenstein:1995un, Hui:2012qt,Sotiriou:2013qea}. 
Therefore, in GR (and many extensions thereof), electro-vacuum black holes are only characterized by their mass, spin, and electric charge. Nevertheless, astrophysical BHs are presumed to carry very negligible charge. Hence, the Kerr metric describes such quiescent black holes. 

That being the case, any deviation from this picture might indicate the existence of new fundamental field(s) or the breakdown of GR. A characteristic example in which BHs admit additional scalar charge is that of a scalar field coupled to the Gauss-Bonnet (GB) invariant $\mathcal{G}=R^{\mu \nu \rho \sigma} R_{\mu \nu \rho \sigma}-4 R^{\mu \nu} R_{\mu \nu}+R^{2}$, where $R_{\mu \nu \rho \sigma}$, $R_{\mu \nu}$, and $R$  being the Riemann tensor, Ricci tensor, and Ricci scalar respectively. The presence of a term of the form $f(\phi) \mathcal{G}$ in the action for some function $f(\phi)$ will lead to a scalar equation of the form
\begin{align}
    \Box \phi + f'(\phi)\mathcal{G} = 0, \quad \Box \coloneqq \nabla_{\mu}\nabla^{\mu}.
\end{align}
In the case of linear coupling, $f(\phi)\propto \phi$, for which the no-hair theorem presented in~\cite{Hui:2012qt} does not hold \cite{Sotiriou:2013qea,Sotiriou:2014pfa}, the only allowed BH solutions are those with a non-trivial-scalar-configuration. On the other hand, assuming stationarity and asymptotic flatness, only GR BHs will be admissible solutions to a theory for which $f'(\phi) = 0$ and $f''(\phi)\mathcal{G}<0$~\cite{Silva:2017uqg}. The first condition is necessary to have $\phi=\rm{constant}$ solutions i.e., GR BHs. Interestingly, if $f''(\phi)\mathcal{G}>0$, the black holes with scalar hair can also appear for a certain mass \cite{Silva:2017uqg,Doneva:2017bvd} or spin \cite{Dima:2020yac,Herdeiro:2020wei,Berti:2020kgk} range. 

To understand the effect of the second condition, consider the linearized version of the scalar equation around $\phi=\phi_{0}$ on a fixed background, then the perturbation $\delta \phi$ will satisfy
\begin{align}
    \left(\Box  + f''(\phi_{0})\mathcal{G}\right)\delta \phi = 0,
\end{align}
from which we can observe that the term $-f''(\phi)\mathcal{G}$ will act as an effective (squared) mass for the perturbation. Hence, $f''(\phi)\mathcal{G}>0$ implies a negative effective mass squared, leading to a tachyonic instability. The scalar then grows exponentially until the non-linear interactions of the theory quench this linear instability. Therefore, the end state of such instability cannot be probed using linear perturbation analysis rather a full non-linear treatment is needed to understand the properties of the scalarized solutions. This phenomenon was first introduced in the context of NSs in~\cite{Damour:1993hw} and called spontaneous scalarization (for a review see~\cite{Doneva:2022ewd}).

The simplest coupling function that will satisfy the previous conditions is quadratic in the scalar field i.e., $f(\phi) \sim \phi^2$ (see also Ref.~\cite{Andreou:2019ikc} for the full set of interactions that contribute to the onset of the instability). The study of static, spherically symmetric BHs with such an interaction was carried out explicitly in Ref.~\cite{Silva:2017uqg} by constructing asymptotically flat scalarized black holes. Later on, it was shown in Ref.~\cite{Dima:2020yac} that rapid rotation can also trigger scalarization. 

One obstacle to exploring the dynamics of these solutions in full generality is the difficulty in obtaining a well-posed initial value problem (IVP) formulation (i.e., the solution should be unique and depend continuously on the initial conditions \cite{hadamard}) for scalar GB theory and, more broadly, for modified theories of gravity. The study of the well-posedness for nonminimal scalars coupled to gravity that are described by the Horndeski action~\cite{Horndeski:1974wa} received much attention recently~\cite{Papallo:2017qvl,Papallo:2017ddx,Ripley:2019hxt,Ripley:2019irj,Ripley:2020vpk,Bezares:2020wkn,Figueras:2020dzx,Figueras:2021abd,Bernard:2019fjb}. It was shown, in the weakly coupled regime where the coupling to the scalar field remains "small", that a well-posed formulation exists~\cite{Kovacs:2020ywu}. Numerical studies restricted to this regime were successfully performed in~\cite{East:2020hgw,East:2021bqk,Corman:2022xqg,AresteSalo:2022hua,AresteSalo:2023mmd}. There is ongoing work on developing a well-posed formulation beyond the weak coupling regime, see {\em e.g.}~Ref.~~\cite{Figueras:2024bba}.

If these theories are understood as nonlinear effective field theories (EFTs), obtained from a more ``fundamental" theory by integrating out heavy fields, then various methods can be employed to tackle the issue of ill-posedness. One method, inspired by viscous relativistic hydrodynamics~\cite{Cayuso:2017iqc}, has been explored in the context of gravity theories in~\cite{Allwright:2018rut,Cayuso:2020lca,Franchini:2022ukz,Cayuso:2023aht,Lara:2021piy,Bezares:2021dma,Bezares:2021yek,Cayuso:2023aht,Gerhardinger:2022bcw,Lara:2024rwa,Corman:2024cdr,Rubio:2024ryv}. Another approach, is to assume that the theory is continuously connected to GR, and treat it perturbatively in the coupling constant that controls the deviations from GR, this way one can generate solutions order by order in the coupling~\cite{Benkel:2016rlz,Benkel:2016kcq,Okounkova:2017yby,Witek:2018dmd,Okounkova:2018abo,Okounkova:2019dfo}. Nevertheless, the application of this method suffers from secular growth~\cite{GalvezGhersi:2021sxs}. It has an inherent limitation in capturing nonlinear effects such as spontaneous scalarization. 

Recently, it was shown in Ref.~\cite{Thaalba:2023fmq} that additional interactions, that might not by themselves introduce new phenomenological effects, can be crucial for well-posedness. This was demonstrated in the case of spherical collapse in scalar GB gravity by the addition of a $\phi^2 R$ interaction.
Interestingly, including this particular interaction helps scalarization models to evade binary pulsar constraints~\cite{Ventagli:2021ubn}, have GR as a cosmological attractor~\cite{Antoniou:2020nax}, and also renders static spherical black holes stable~\cite{Antoniou:2022agj}. 

Ref.~\cite{Thaalba:2023fmq} considered the collapse of a scalar cloud. Here, we use a different numerical implementation that employs excision to expand on those results and consider a broader set of interesting scenarios for spherical black hole dynamics. In section~\ref{sec:theory}, we present details of scalar-GB-Ricci gravity and its equations of motion in the chosen gauge. In section~\ref{sec:setup} we describe the numerical methods upon which the code is built. In section \ref{subsec:Well-posedness} we briefly revisit the effect of the Ricci coupling on well-posedness, focusing on coupling parameters for which the final BH is scalarized. We observe that excision leads to well-posed evolution for a larger part of the parameter space, as the ill-posed region was beyond the horizon in certain cases.
In section \ref{subsec:min_mass} we study the dynamics of black holes that are near the threshold of scalarization or near the minimum mass threshold, which is a characteristic feature of black holes in this class of theories. We also study initial data for which the initial black hole lies near but below the minimum mass threshold.
In section~\ref{subsec:QNM} we extract the monopolar QNMs, while in section~\ref{subsec:tach_ins} we estimate the time scale of the tachyonic instability that leads to scalarization.

\section{The Theory}\label{sec:theory}
We consider the following action: 
\begin{align}
    \label{eq:action}
    S= \frac{1}{16 \pi} \int \mathrm{d}^{4} x \sqrt{-g}\left[R+X- \left(\frac{\beta}{2}R - \alpha \mathcal{G}\right)f(\phi)\right],
\end{align}
where $g = \text{det}(g_{\mu \nu})$, $X=-\nabla_\mu\phi\nabla^\mu\phi/2$. We use units of $G=c=1$. The parameter $\beta$ is a dimensionless coupling constant while $\alpha$ is dimension length squared.

Varying action \eqref{eq:action} with respect to the metric yields,
\begin{align}
    \label{eq:einstein}
    G_{\mu \nu} &= T^{\phi}_{\mu \nu}~,
\end{align}
where
\begin{align}
    T_{\mu \nu}^{\phi}&= -\frac{1}{4} g_{\mu \nu}(\nabla \phi)^{2}+\frac{1}{2} \nabla_{\mu} \phi \nabla_{\nu} \phi \nonumber \\  
    & -\frac{\alpha}{g} g_{\mu(\rho} g_{\sigma) \nu} \epsilon^{\kappa \rho \alpha \beta} \epsilon^{\sigma \gamma \lambda \tau} R_{\lambda \tau \alpha \beta} \nabla_{\gamma} \nabla_{\kappa} f(\phi) \nonumber \\ 
    & +\frac{\beta}{2}\left(G_{\mu \nu} + g_{\mu \nu} \Box - \nabla_{\mu}\nabla_{\nu}\right)f(\phi)~.
    \label{eq:scalar_stressenergy}
\end{align}
and with respect to the scalar fields, gives
\begin{align}
    \label{eq:scalar}
    \Box \phi &= \left(\frac{\beta}{2}R -\alpha\mathcal{G}\right)f^{\prime}(\phi)~,
\end{align}
where $G_{\mu\nu}$ is the Einstein tensor, and $\epsilon^{\sigma \gamma \lambda \tau}$ is the Levi-civita totally anti-symmetric tensor density. In what follows, we restrict our analysis to the $f(\phi) = \frac{\phi^2}{2}$ model.

We choose to work with Painleve-Gullstrand (PG) like coordinates in spherical symmetry~\cite{PhysRevD.79.101503,Kanai:2010ae,Ripley:2019aqj}. These coordinates are horizon penetrating, i.e., the metric functions remain regular through the formation of an apparent horizon. Consequently, the genesis and the horizon dynamics can be followed without encountering the singularity present in Schwarzschild coordinates.
The line element is given by 
\begin{align}
    \mathrm{d} s^2&=-A(t, r)^2 \mathrm{d} t^2+\left[\mathrm{d} r+A(t, r) \zeta(t, r) \mathrm{d} t\right]^2 \nonumber \\ 
    &+r^2\left(\mathrm{d} \theta^2+\sin ^2 \theta \mathrm{d} \varphi^2\right),
\end{align}
and the scalar field is also spherically symmetric, i.e., $\phi=\phi(t,r)$. To write the  equations of motion (EoM) as a first-order system of partial differential equations (PDEs), we introduce the following variables 
\begin{align}
    Q & \equiv \partial_{r}\phi,\\
    P & \equiv \frac{1}{A} \partial_{t}\phi - \zeta Q.
\end{align}
Then, the evolution equations for $\phi$ and $Q$ can be obtained from the definitions of $P$ and $Q$ as 
\begin{align}
    E_{\phi} & \equiv \partial_{t} \phi - A(P+\zeta Q) = 0, \\ 
    E_{Q} & \equiv \partial_{t} Q - \partial_{r}\left(A(P+\zeta Q)\right) = 0,
\end{align}
and, from the EoM we obtain evolution equations for $P$ and $\zeta$, and constraint equations for $\zeta$ and 
$A$, which can be written schematically as 
\begin{align*}
    E_{P} & \equiv E_{P}(\partial_{t}P; A, \zeta, \phi, P, \partial_{r}P, Q, \partial_{r}Q)=0, \\
    E_{\zeta} & \equiv E_{\zeta}(\partial_{t}\zeta; \zeta, \partial_{r}\zeta, A, \partial_{r}A, \phi, P, \partial_{r}P, Q, \partial_{r}Q)=0, \\
    C_{A} & \equiv C_{A}(\partial_{r}A; \zeta, \phi, P, \partial_{r}P, Q, \partial_{r}Q)=0, \label{C_alpha}\\  
    C_{\zeta} & \equiv C_{\zeta}(\partial_{r}\zeta; A, \phi, P, \partial_{r}P, Q, \partial_{r}Q)=0.
\end{align*}

\section{Setup}\label{sec:setup}
In this section, we discuss the numerical implementation and the various techniques employed to evolve BHs with scalar perturbation. We then describe the initial data used to numerically study the dynamics of BHs in the theory under consideration. We further elaborate on useful diagnostic tools that allow us to keep track of the hyperbolic nature of the equations. Additionally, we present a brief description of static BH solutions. Finally, we present our results.  

\subsection{Numerical methods}
We discretize the numerical domain for a given resolution $\rm{N}$ non-uniformly. This is done by employing the so-called fish-eye coordinate~\cite{Baker:2001sf,Alcubierre:2002kk,Corelli:2022phw}. This allows for more resolution in high curvature regions where the derivatives of the scalar are very steep and changing rapidly. In this setup, we define our areal radial coordinate $r$ as a function of a uniform grid $x$ as follows  
\begin{align}
    r(x) &= \eta_2 x + \left( 1 - \eta_1 \right) \log\left( \frac{1+\mathrm{e}^{x_1-x}}{1+\mathrm{e}^{x_1}} \right) \nonumber \\ 
    &- \left( 1 - \eta_2 \right) \log\left( \frac{1+\mathrm{e}^{x_2-x}}{1+\mathrm{e}^{x_2}} \right)~,
\end{align}
where $\eta_1\leq 1$, $\eta_2 \geq 1$ and $0<x_{1}<x_{2}$ are real parameters.
By definition, the spatial step of the uniform grid is $\Delta x = X/\rm{N},$ where $X$ denotes the outer boundary. The time step is given by $\Delta t = \lambda \Delta x$ where $\lambda$ is the Courant factor. The uniform spatial derivatives are discretized using a second-order finite differences operator satisfying summation by parts~\cite{Calabrese:2003vx}. Since we solve the equations in the areal radius grid, we need to transform the radial derivatives with the Jacobian
\begin{equation}
    \frac{\partial f}{\partial r} = \frac{\partial x}{\partial r} \frac{\partial f}{\partial x}~.
\end{equation}
During the evolution, we keep track of the expansion of null congruences, which determine the location of the apparent horizon. In these coordinates, the expansion is proportional to $1-\zeta$; therefore, the horizon is at $\zeta=1$, and we update its location accordingly. Nevertheless, we only perform this update if the horizon increases in size. After pinpointing the horizon, we place the excision surface $r_{\text{exc}}$ inside of it.
Furthermore,  we chose $\eta_{1} =0.25,$ $\eta_{2} = 15,$ $x_1=51$, and $x_{2}=100.$ This choice ensures that the sampling of the areal radius $r$ is more dense close to the excision surface and more sparse towards the last point $R = r(X)$. Given the initial data at $t=0$, we solve the constraint equations for $\zeta$ and $A$ on $[r_{\text{exc}},R]$ using a second-order Runge Kutta (RK2) scheme. 

We employ a one-sided stencil to compute radial derivatives at the excision surface. We solve a time evolution equation for $\phi, P, Q$, and $\zeta$, by using the method of lines for which we discretise radial derivatives using second-order finite differences, and integrate in time using a fourth-order Runge Kutta and a fourth-order Kreiss-Olliger (KO) dissipation term. On the other hand, the metric function $A$ is obtained by solving its constraint equation Eq.~\eqref{C_alpha} by choosing $A(t,r_{\text{exc}}) = 1$ and then, using residual gauge freedom, we rescale $A$ such that $A(t,R)=1$. We have followed a similar procedure to the one discussed in~\cite{Ripley:2019aqj}.

\subsection{Initial data}
We evolve a scalar field on a black hole background with mass $M_{\text{BH}}$. The region inside the apparent horizon, located at $\zeta=1$, is excised. At the excision radius, the shift $\zeta$ and the lapse $A$ are set to their GR values, namely
\begin{align}
    \zeta(0, r_{\text{exc}}) &= \sqrt{\frac{2M_{\text{BH}}}{r_{\text{exc}}}}, \\ 
    A(0, r_{\text{exc}}) &= 1.
\end{align}
The scalar field $\phi$ and its time derivative are given by
\begin{align}
\label{eq:initialdata2}
    \phi(0,r) &= a_0 \left(\frac{r}{w_{0}}\right)^2\exp\left[-\left(\frac{r-r_0}{w_0}\right)^2\right],\\
    P(0,r) &= \frac{1}{r}\phi(0,r) + Q(0,r),
\end{align}
where $a_{0}$, $r_{0}$ and $w_0$ are constants. We fix $M_{\text{BH}}=1$, $r_0 = 25$, and $w_0 = 6$ unless otherwise stated. In subsection \ref{subsec:QNM}, we will use the following initial data field
\begin{align}
     \phi(0,r) &= a_{0}r\exp\left[-\left(\frac{r-r_0}{w_0}\right)^2\right], \\
     P(0,r) &= 0,
\end{align}
with $a_{0} = 0.01, r_0 = 10$, and $w_0 = 2$, for studying the quasi-normal modes (QNMs).

\subsection{Well-posedness and diagnostics}
Generally, a PDE system is considered well-posed if it admits a unique solution that depends continuously on the initial data \cite{hadamard}. Moreover, an initial value (Cauchy) problem is strongly hyperbolic if the principal part (the highest derivative terms) is diagonalizable with real eigenvalues. It can be proven that such a property is sufficient for well-posedness~\cite{Hilditch:2013sba}. These eigenvalues are the phase velocities, also called characteristic speeds, of the system's plane wave solutions.
As such, the speeds represent a useful diagnostic tool in numerical considerations. In spherical symmetry, they are given by~\cite{Ripley:2019irj,Ripley:2019aqj}
\begin{align}
    \label{eq:char_speeds}
    c_{\pm} = \frac{1}{2} \left(\text{Tr}(\mathcal{C}) \pm \sqrt{\mathcal{D}}\right),
\end{align}
where $\mathcal{C}$ and $\mathcal{D}$ depend on the functions $A$, $\zeta$, $\phi$, $P$, and $Q$, and their derivation can be found in Appendix~\ref{appendix:I}. We say that the system is hyperbolic if the characteristic speeds are real ($\mathcal{D}>0$), elliptic if they are complex ($\mathcal{D}<0$), and parabolic if the speeds are degenerate ($\mathcal{D}=0$). 
We keep track of the discriminant to ensure that the PDEs are hyperbolic during the evolution. 
We excise the elliptic region if it is confined to the apparent horizon. If the elliptic region emerges or appears outside the horizon, then we stop the evolution. We also compute the Misner Sharp mass as $M(t,r) \equiv \frac{r}{2}\left(1-(\nabla r)^2\right) = \frac{r}{2}\zeta(t,r)^2$~\cite{PhysRev.136.B571}. To capture the spacetime mass denoted as $M,$ we evaluate $M(t,r)$ at the outer boundary of the numerical domain. 

\subsection{Static BH solutions}
A characteristic feature of black holes with scalar hair in theories that include an interaction between the Gauss-Bonnet invariant and a scalar field is that their scalar charge is not a free parameter, but it is instead uniquely determined by their mass (and spin for rotating black holes) ~\cite{Kanti:1995vq,Sotiriou:2014pfa, Antoniou:2021zoy,Thaalba:2022bnt}. The relation between the scalar charge and the mass is dictated by a regularity condition on the horizon.

\begin{figure}
     \centering
    \includegraphics[width=1\linewidth]{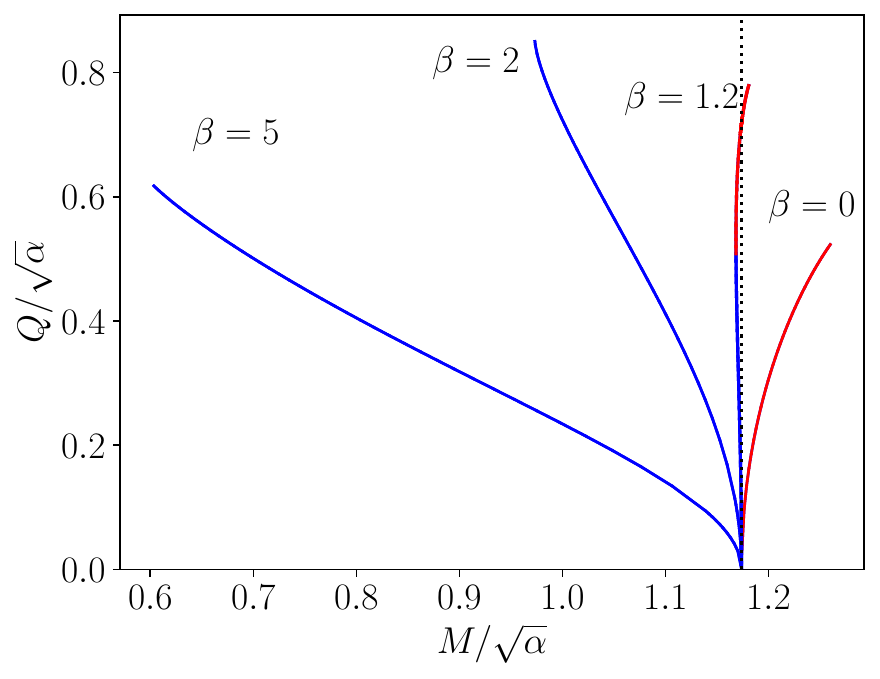}
     \caption{Charge-Mass plot for scalarized BHs. The blue curves correspond to stable solutions while the red to unstable ones as shown in \cite{Antoniou:2022agj}. The threshold mass, marked by the black dotted line, is $M_{\text{th}}/\sqrt{\alpha} \approx 1.174$, while the minimum mass varies for different values of $\beta$.}
     \label{fig:QM_1}
\end{figure}

Static, spherically symmetric black holes in the theory we consider here were first studied in~\cite{Antoniou:2021zoy,Antoniou:2022agj}. Black hole solutions with a non-trivial scalar profile, dubbed scalarized BHs, only exist for a specific range of masses, i.e., only for $M \in [M_\text{min},M_{\text{th}}]$. Black holes with mass $M < M_{\text{min}}$ do not exist (as already discussed), while if $M > M_{\text{th}}$, the only stable black holes are those of GR. Figure~\ref{fig:QM_1} shows the relation between the scalar charge and the mass of such scalarized BHs. To find these solutions we follow a similar procedure to~\cite{Antoniou:2021zoy}. 

\section{Results}

\subsection{Well-posedness analysis}\label{subsec:Well-posedness}

First, we briefly examine the effect of the $\phi^2 R$ term on well-posedness. This problem was considered in~\cite{Thaalba:2023fmq} and our results are in agreement with the earlier findings, namely that, for a given $\alpha/M^2$, large enough values of the coupling constant $\beta$ render the system hyperbolic. Here we have focused on a choice of parameters for which black holes are scalarized. This case is more interesting because scalarization is a non-linear process and one could expect to encounter large gradients, which are prone to cause loss of hyperbolicity. Therefore, examining the effect of the Ricci term in this regime is particularly interesting. These results are displayed in table~\ref{tab:runs}. The table shows the outcome of different simulations (either loss of hyperbolicity or a scalarized BH) for various coupling constants $\alpha/M^2$, and $\beta$.

\begin{figure}
     \centering
    \includegraphics[width=1\linewidth]{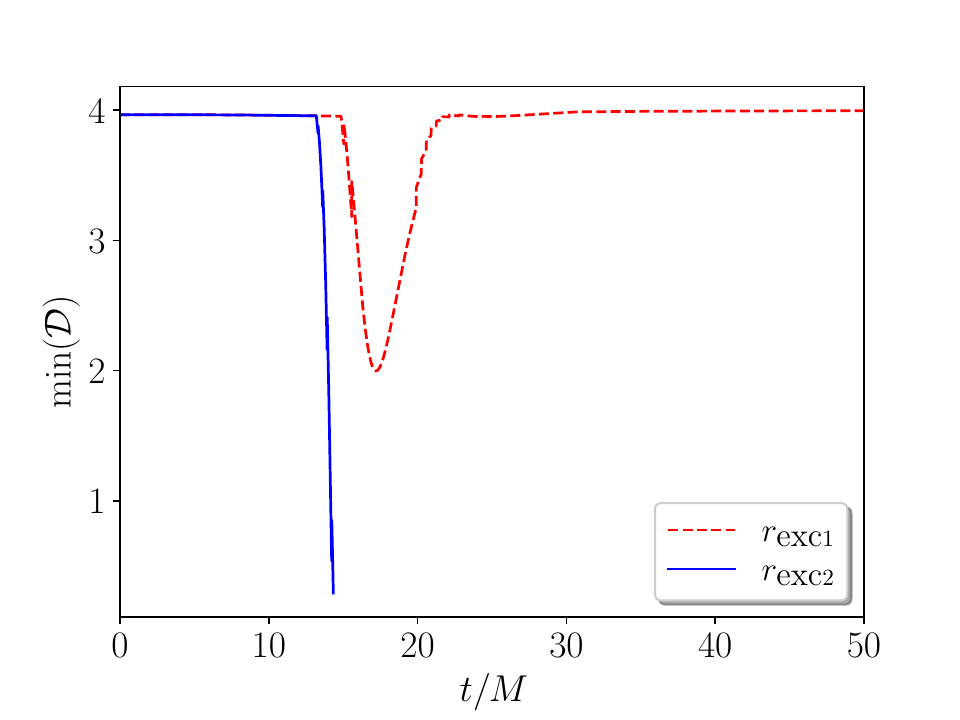}
     \caption{The minimum value of the discriminant $\mathcal{D}$ in the radial direction as a function of time for two different excision locations $r_{\text{exc}1},r_{\text{exc}2}$ such that $r_{\text{exc}1}$ is closer to the trapped surface from the interior. In this case, $\alpha/M^2=1$, $\beta=2$ with a scalar pulse of amplitude $a_{0}=2 \times 10^{-3}$. The blue curve corresponds to a fixed location of the excision surface and this eventually leads to loss of hyperbolicity. On the other hand, when the excision surface location is allowed to move closer to the trapped surface from the interior the elliptic region is excised (red dashed curve), and the discriminant remains positive throughout the evolution.}
     \label{fig:ER_Ex}
\end{figure}

An interesting feature we observe is that, when the equations only change character in a region inside the BH, then excising this region enlarges the parameter space for which the system is well-posed~\cite{Ripley:2019aqj}.  We illustrate this fact in Fig.~\ref{fig:ER_Ex}. In this case, the excision surface is allowed to move closer to the horizon, thereby removing the elliptic region from the numerical domain. Hence, the evolution will proceed without impediment due to ill-posedness inside the horizon. This type of excision leaves the physics of the exterior unaffected since the region inside the BH is causally disconnected from the exterior.
\begin{table}
    \centering
    \begin{tabular}{cc|c|c}
       \multicolumn{2}{c|}{Coupling constants}&\multirow{2}*{Outcome}\\
        $\alpha/M^2$ & $\beta$ &\\
       \hline
          0.75& 0.0 &  LoH\\
          0.75& 0.5 &  LoH\\
          0.75& 2.0 &  sBH\\
          1.25& 0.0 &  LoH\\
          1.25& 2.0 &  LoH\\
          1.25& 3.5 &  sBH\\
          1.75& 0.0 &  LoH\\
          1.75& 3.0 &  LoH\\
          1.75& 4.0 &  sBH\\
          2.25& 0.0 &  LoH\\
          2.25& 5.0 &  LoH\\
          2.25& 5.5 &  sBH
    \end{tabular}
    \caption{The initial scalar pulse has $a_{0}=2\times10^{-3}$. The initial Misner-Sharp mass is indicated by $M$. The outcome indicates the end state of the evolution. The outcome is either loss of hyperbolicity (LoH), or a scalarized black hole (sBH). It is evident, in accordance with~\cite{Thaalba:2023fmq}, that the addition of the Ricci coupling alleviates the LoH problem for certain choices of $\beta$.}
    \label{tab:runs}
\end{table}

\subsection{Dynamics of Minimum and threshold mass BHs}
\label{subsec:min_mass} 
Next, we explore the dynamical behaviour of the BH solutions near the minimum and threshold mass through non-linear simulations. To this end, we consider different setups. 

\begin{figure}
     \centering
    \includegraphics[width=1\linewidth]{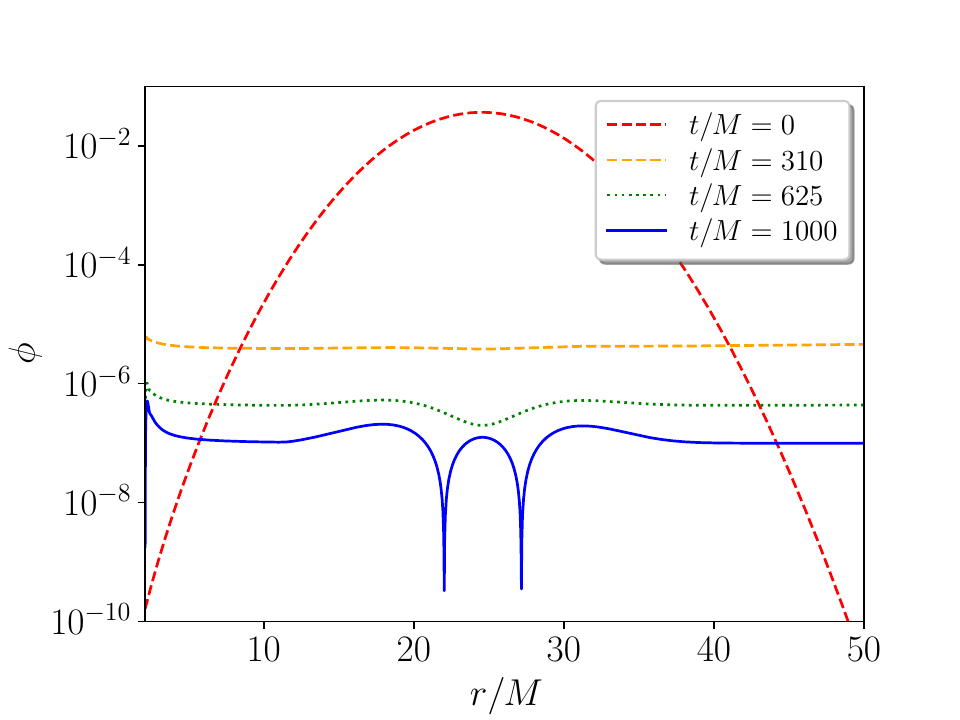}
     \caption{Scalar field profile at different times for $\alpha/M^2=0.25$, $\beta = 0.5$ with $a_0 = 2\times10^{-3}$. This choice of parameters corresponds to a setup where the initial $M>M_{\text{th}}$ i.e., this would be a hairless black hole.}
     \label{fig:Sch}
\end{figure}

\begin{figure}
     \centering
    \includegraphics[width=1\linewidth]{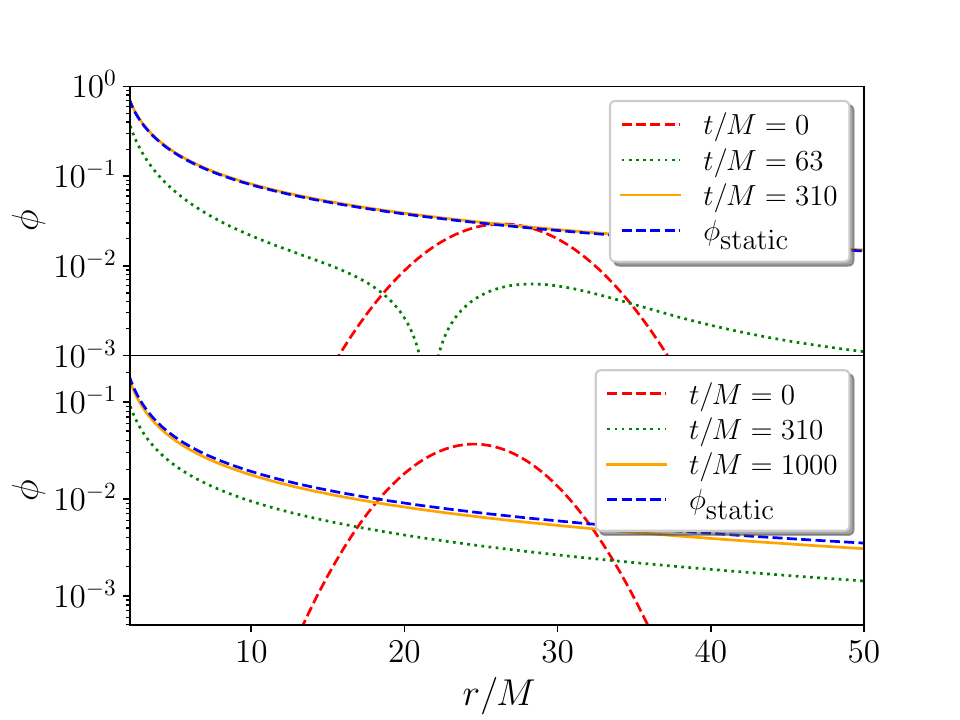}
    \caption{Scalar profile for different times. The static solution is indicated by $\phi_{\text{static}}$. The top panel is for $\alpha/M^2=0.75$, $\beta = 2$, with an initial scalar profile described by $a_0 = 2\times10^{-3}$. In the bottom panel we have $\alpha/M^2=1, \beta=2$, with $M_{\text{BH}}=0.95$, and initial scalar profile with $a_0 = 1.6\times10^{-3}$. The mass of the black hole (only without the scalar field contribution) in this setup is smaller than the minimum mass.}
    \label{fig:phi}
\end{figure}

We first consider initial data such that the initial mass $M$ of the system is larger than $M_{\text{th}}$. The end state of the evolution is an apparent horizon with a negligible scalar field, consistent with the stationary analysis~\cite{Antoniou:2022agj,Antoniou:2021zoy}. This is illustrated in Fig.~\ref{fig:Sch} where the late time behaviour of the scalar field approaches zero.
We then consider a case where the initial mass is below $M_{\text{th}}$ and the initial black hole mass in larger than the minimum mass $M_{\text{min}}$. As expected, the end state is a black hole with a nontrivial scalar profile, see top panel of Fig~\ref{fig:phi}. We also consider an initial configuration in which the total mass is below $M_{\text{th}}$ while the initial black hole mass is smaller than the minimum mass $M_{\text{min}}$. In this case, we choose the scalar pulse such that it contributes enough mass to the system to have $M_{\text{min}}<M<M_{{\text{th}}}$. Interestingly, the evolution progresses smoothly and the endpoint is again a scalarized black hole, as depicted in the bottom panel of Fig~\ref{fig:phi}.

\begin{figure}
     \centering
    \includegraphics[width=1\linewidth]{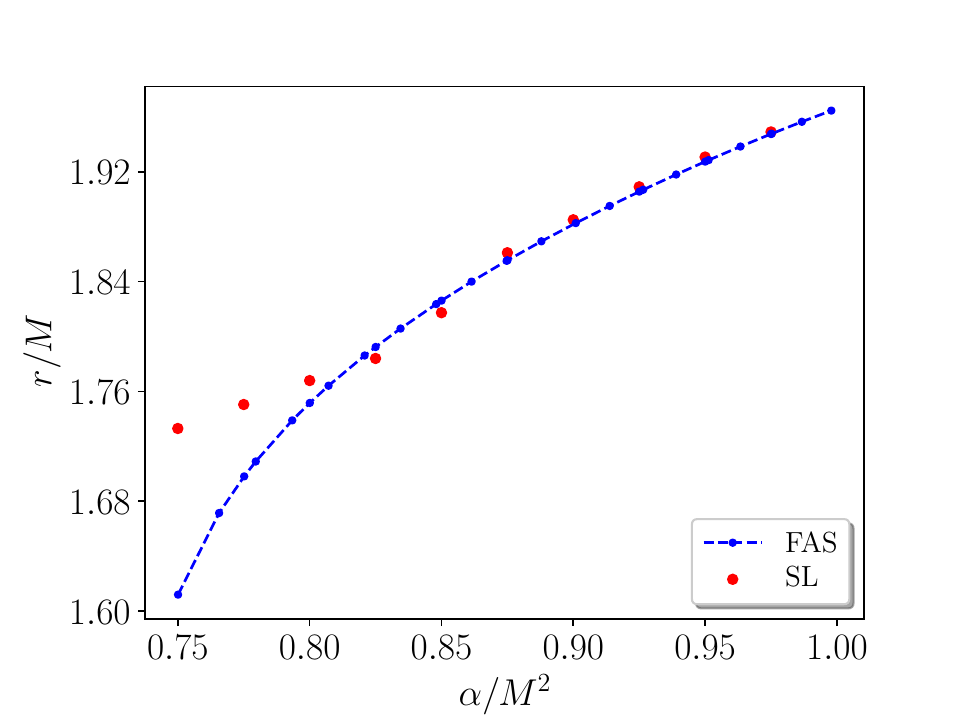}
     \caption{A comparison between the finite area singularity (FAS) radius (in dashed blue) extracted from the static solution and the size of the elliptic region (in red dots) formed inside the horizon, which is excised along the sonic line (SL). This is for $\beta=2$.}
     \label{fig:FAS}
\end{figure}

Finally, to examine a possible connection between the finite area singularity and an elliptic region forming inside the apparent horizon, we excise the elliptic region along the sonic line and compare the size of the elliptic region with that of the finite area singularity of the static solution as in Fig~\ref{fig:FAS}. We observe a good agreement between both quantities especially for smaller BHs.

\subsection{Quasi-Normal modes}\label{subsec:QNM}
\begin{figure*}
     \centering
    \includegraphics[ width=\linewidth]{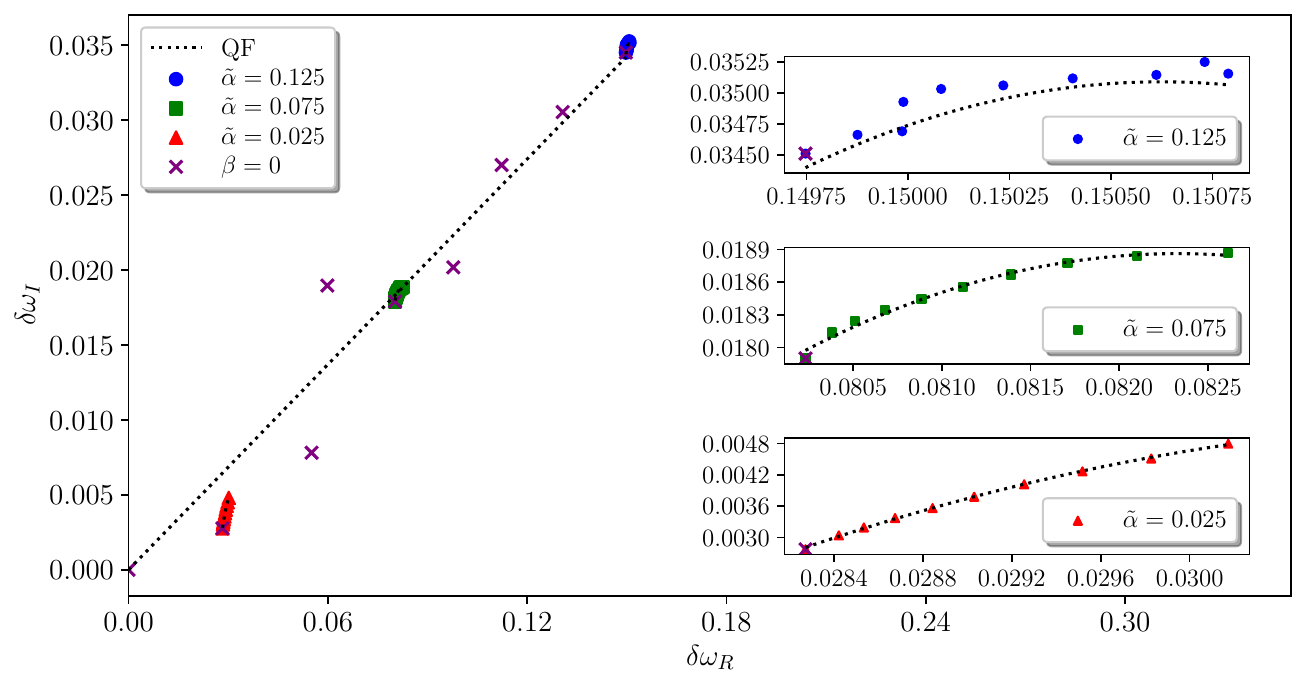}
     \caption{Relation between the normalized relative difference of the real and imaginary parts of the QNMs for different values of $\tilde{\alpha} = \alpha/M^2$ and a quadratic fit (QF) thereof. As the value of $\beta = \{0.5,0.75,1,1.25,1.5,1.75,2,2.25,2.5\}$ increases the deviation in the real and imaginary parts from their GR values grows.}
      \label{fig:delta_omega}
\end{figure*}

\begin{figure}
     \centering
    \includegraphics[width=1\linewidth]{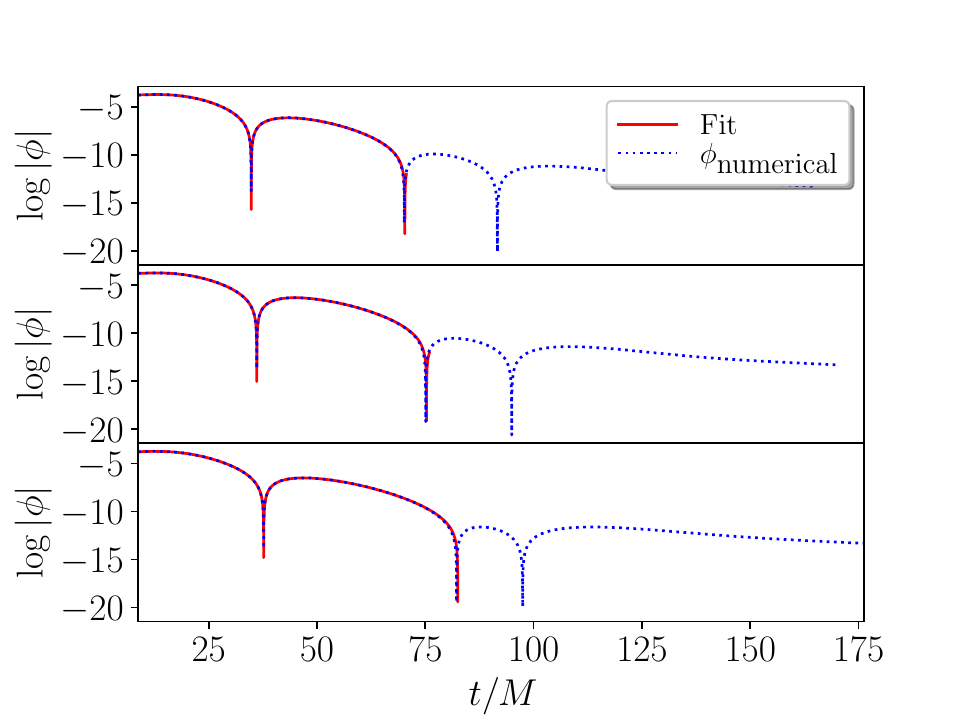}
     \caption{The scalar field extracted from the numerical simulation (blue dotted line) and the damped-sinusoid fit (red solid line). Each plot is done for the case $\beta=2.25$ and, from top to bottom, $\alpha/M^2 = 0.025, 0.075, 0.125$. }
     \label{fig:QNMs}
\end{figure}

Probing the geometry of a BH using test (scalar, vector, or tensor) fields by utilizing linear perturbation theory proved very powerful~\cite{Kokkotas:1999bd}. Due to the nature of a BH where the horizon acts as a dissipation in the system, the behaviour of these fields as a perturbation is described by exponentially decaying oscillations; the so-called quasi-normal modes. The rate of decay and the oscillation frequency are related to the mass, and spin of the BH. In spherical symmetry, there are no tensor propagating degrees of freedom. We can only extract the $\ell=m=0$ scalar QNMs of BH ringing, by fitting the extracted value of the scalar field at $r_\text{ext} = 3 r_{0}$ as 
\begin{align}
    \phi(t) = \sum_{n=0}^{\infty} A_{n}\exp(-\omega_{I,n}t)\sin(\omega_{R,n}t + \varphi_{n}),
\end{align}
where $n$ denotes the overtone index. The quantities $A_n$, $\omega_{n} = \omega_{R,n}-i\omega_{I,n}$, and $\varphi_{n}$ are the mode's amplitude, complex frequency, and phase respectively. To obtain an accurate fit of the decaying oscillatory behaviour it helps to have as many oscillations as possible before the appearance of the tail. In the cases we considered, we had to tweak the initial data to achieve a few oscillations to allow for an accurate fit. Such fits for different values of $\alpha/M^2$ are shown in Fig.~\ref{fig:QNMs}. The tail behaviour appears very quickly for larger values of $\alpha/M^2$, which hinders the accurate fitting of the QNMs. Due to this, we only considered lower values of $\alpha/M^2$, for which the background solution is given by Schwarzschild metric, according to the analysis in~\cite{Antoniou:2021zoy, Antoniou:2022agj}.  Nonetheless, we observe a deviation from the GR QNMs since the Gauss-Bonnet invariant is non-trivial for the Schwarzschild metric.  The results for 
\begin{equation}
    \delta \omega_{R} = \left| \frac{\text{Re}(\omega_{\text{GR}}-\omega)}{\text{Re}\,\omega_{\text{GR}}} \right|, \quad \delta \omega_{I} = \left| \frac{ \text{Im}(\omega_{\text{GR}}-\omega)}{\text{Im}\, \omega_{\text{GR}}} \right|
\end{equation}
are summarised in Table~\ref{tab:QNMs} and Figure~\ref{fig:delta_omega}. 
Fitting the deviation in the QNMs from GR for $\beta=0$, we have
\begin{equation}
    \delta \omega_{(\alpha,0)} = \sum_{k=0}^2 c_k \left(\frac{\alpha}{M^2}\right)^k \,,
\end{equation}
where the coefficients $c_k$ are given in table~\ref{tab:QNM_fit}.
If we turn on the $\beta$ term and fit the deviation that it causes to the QNMs from the GB theory (i.e., $\alpha \ne 0, \beta=0$), we obtain \begin{equation}
    \delta \omega_{(\alpha,\beta)} = \delta \omega_{(\alpha,0)} + \sum_{k=1}^2 d_k(\alpha) \beta^k \,,
\end{equation}
where the coefficients $d_k(\alpha)$ are given in Table~\ref{tab:QNM_fit} for the values $\alpha = [0.025, 0.075, 0.125]$.

\begin{table}
    \centering
    \begin{tabular}{c|cc}
        & $\mathrm{Re}$ & $\mathrm{Im}$ \\
       \hline
       $\alpha = 0$ & & \\
       \hline
       $c_0$ & $-1.0539\times 10^{-3}$ & $1.99309\times 10^{-3}$ \\
       $c_1$ & $2.01256\times 10^{-1}$ & $8.59663\times 10^{-1}$ \\
       $c_2$ & $6.9759\times 10^{-1}$ & $2.52698$ \\
       \hline
       $\alpha = 0.025$ & & \\
       \hline
       $d_1$ & $1.56327\times 10^{-4}$ & $4.61635\times 10^{-4}$ \\
       $d_2$ & $2.47143\times 10^{-4}$ & $1.40432\times 10^{-4}$ \\
       \hline
       $\alpha = 0.075$ & & \\
       \hline
       $d_1$ & $3.63718\times 10^{-4}$ & $3.85884\times 10^{-5}$ \\
       $d_2$ & $1.00527\times 10^{-4}$ & $5.0495\times 10^{-4}$ \\
       \hline
       $\alpha = 0.125$ & & \\
       \hline
       $d_1$ & $2.29461\times 10^{-4}$ & $4.50753\times 10^{-4}$ \\
       $d_2$ & $1.15081\times 10^{-4}$ & $5.90922\times 10^{-5}$ \\
    \end{tabular}
    \caption{The real and imaginary part of the quadratic fit coefficients for quasi-normal modes for different values of the couplings $\alpha/M^2$.}
    \label{tab:QNM_fit}
\end{table}

\subsection{Tachyonic instability}
\label{subsec:tach_ins}
Lastly, we examine the time scale of the tachyonic instability that is responsible for the existence of scalarized black holes. In simple terms, linear analysis has shown that the tachyonic instability causes the scalar field to grow exponentially \cite{Silva:2017uqg,Doneva:2017bvd},  see Ref.~\cite{Antoniou:2021zoy} for an analysis specific to the model we consider here and Ref.~\cite{Doneva:2022ewd} for a review. Nevertheless, the end state of the instability is controlled by the non-linear interactions present in the theory \cite{Silva:2018qhn,Macedo:2019sem,Antoniou:2021zoy}. In the simulations, we observe both the linear instability and its nonlinear quenching. We are also able to estimate the timescale of the instability. 
To monitor the goodness of the results, we compare the numerical extraction against an analysis in the static limit assuming linear perturbations with a continued fraction method. The details are explained in appendix~\ref{app:linear_perturbation}.

\begin{figure}
     \centering
    \includegraphics[width=1\linewidth]{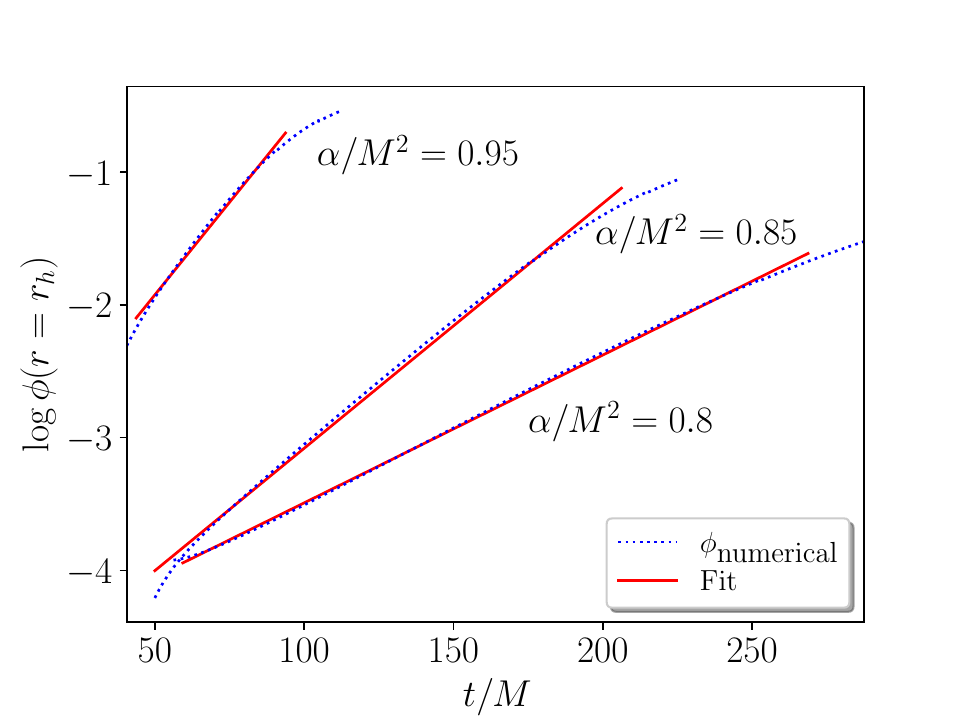}
     \caption{The growth of the scalar field at the horizon of a BH due to the tachyonic instability which is then quenched. The dashed red line is a linear fit to the exponential growth of the scalar field (in log scale). The slope of the fit is the inverse of the time scale of the instability. This is for $\beta = 2$, and $a_{0} = 2 \times 10^{-3}$.}
     \label{fig:tau}
\end{figure}

The behaviour of the exponential growth of the scalar field at the horizon (in log scale) is shown in Fig.~\ref{fig:tau} with a linear fit from which we extract the time scale (see table~\ref{tab:time_scale} for more cases and the results from the linear perturbation analysis). At larger times, the exponential growth clearly stops. 

\begin{table}
    \centering
    \begin{tabular}{cc|cc}
        \multicolumn{2}{c|}{Coupling constants} & \multirow{2}*{$1/\tau^\mathrm{fit}$} & \multirow{2}*{$1/\tau^\mathrm{cf}$} \\
         $\alpha/M^2$ & $\beta$ & & \\
       \hline
         0.750 & 2.0 & 0.00353 & 0.004305 \\
         0.775 & 2.0 & 0.00713 & 0.008509 \\
         0.800 & 2.0 & 0.0111 & 0.01254 \\
         0.825 & 2.0 & 0.0151 & 0.01643 \\
         0.850 & 2.0 & 0.0184 & 0.02019 \\
         0.875 & 2.0 & 0.0210 & 0.02383 \\
         0.900 & 2.0 & 0.0236 & 0.02737 \\
         0.925 & 2.0 & 0.0259 & 0.03082 \\
         0.950 & 2.0 & 0.0279 & 0.03418 \\
         0.975 & 2.0 & 0.0317 & 0.03746 \\
         1.0 & 2.0 & 0.0337 & 0.04067 \\
    \end{tabular}
    \caption{The inverse of the tachyonic instability time scale. The first column reports the values extracted with a fit from the numerical solution, while the second column reports the timescales evaluated with a continued fraction method.}
    \label{tab:time_scale}
\end{table}

\section{Conclusions}
We have explored the dynamics of spherically symmetric black holes in scalar Gauss-Bonnet gravity with an additional Ricci coupling. This additional coupling was shown to help mitigate the ill-posed behaviour of the equations in spherical symmetry \cite{Thaalba:2023fmq} and there is a positive indication of a similar effect in $3+1$ simulations \cite{Doneva:2024ntw}. We observe similar behaviour for black hole initial data. We also show that the part of the parameter space where spherical evolution remains well-possed can be enlarged by excising a region inside the horizon in which hyperbolicity is lost. Since this can be done without affecting the physics of the exterior, it can be seen as an advantage of excision methods in such setups \cite{Ripley:2019aqj}.

We have considered initial data that are expected to lead to black holes of different masses. The cases in which the initial mass is larger than the scalarization threshold mass led to a negligible scalar field at late times,  consistent with static analysis. For initial data within the mass range for which scalarization occurs, we have successfully matched the late-time scalar field profile obtained from the simulation with the one obtained by solving the static equations of motion. Finally, we have considered initial data for which the initial black hole mass was below the minimum mass threshold of the theory, but the initial scalar pulse contributed enough mass so that the final black hole was expected to be above the minimum mass threshold. We did not encounter any issues in evolving the system and capturing the formation of a scalarized black hole.

The presence of a finite-area singularity in the interior of scalarized black holes is poorly understood. Here, we explored the relation between the appearance of such a singularity and the loss of hyperbolicity in black hole spacetimes. This is motivated by the relation between the minimum mass and the finite-area singularity. That is, the minimum mass is approached as the finite-area singularity grows. We found a good agreement between the radius of the finite-area singularity and the location of the sonic line. 

Additionally, we have extracted the monopolar scalar QNMs for a range of couplings and note a quadratic behaviour for the deviations, in the real and imaginary parts of the modes, from GR. Finally, we extracted the timescale of the linear tachyonic instability associated with scalarization from our numerical simulations,  by fitting the linear growth of the scalar field before it is eventually quenched. To check the numerical results for the QNMs and the time scale of the instability, we produced a continued fraction analysis of linear perturbations in the static limit.

It would be interesting to generalize our study to the case of spinning black holes using $3+1$ simulations and to also consider the more realistic scenario of stellar collapse. One would expect that the dynamics of the finite-area singularity would be more intricate in spinning black holes. Hence, a credible relation with the loss of hyperbolicity might be less straightforward and more challenging to establish. Additionally, one could try to assess numerically the non-linear stability of scalarized black holes by studying more general initial data. Determining whether the loss of hyperbolicity and the positive effect of the Ricci coupling is due to the gauge choice or the propagating physical degrees of freedom is already underway by utilising the results of~\cite{Reall:2021voz}.

\acknowledgements
We thank Ramiro Cayuso, Fabrizio Corelli, Guillermo Lara and Sebastian Volkel for useful discussions.
TS acknowledge partial support from the STFC Consolidated Grant nos. ST/V005596/1 and ST/X000672/1. MB acknowledge partial support from the STFC Consolidated Grant no. ST/Z000424/1

\appendix

\section{Characteristic speeds}\label{appendix:I}
As mentioned in the main text, the speeds deduced from a PDE system indicate its character. To see this, consider a first-order system $V_{I}(x, u, \partial u)=0$, with the index $I$ counting the number of equations, $x^{\mu}$ being the spacetime coordinates, and $u^{J}$ denoting the field content. The principal symbol is defined as follows \cite{Ripley:2022cdh,Ripley:2019irj,Kovacs:2020ywu,Sarbach:2012pr}
\begin{align}
    \mathcal{P}_{IJ}(\xi) \coloneqq \mathcal{P}_{IJ}^{\mu} \xi_{\mu}= \frac{\partial V_{I}}{\partial(\partial_{\mu}u^J)}\xi_{\mu},
\end{align} 
for some covector $\xi_{\mu}$. If the PDE system is describing a Cauchy problem then a necessary condition for the system to be well-posed is for the solutions of the characteristic equation 
\begin{align}
    \label{eq:char_eqn}
    \text{det}(\mathcal{P}_{IJ}(\xi)) = 0.
\end{align}
to be real and distinct. In spherical symmetry, the principal symbol matrix is given by \cite{Ripley:2022cdh}.
\begin{align}
    \begin{aligned}
        \mathcal{P}(\xi)=\left(\begin{array}{cc}
        \mathcal{A} \xi_t+\mathcal{B} \xi_r & \mathcal{Q} \xi_r \\
        \mathcal{R} \xi_r & \mathcal{S} \xi_r
        \end{array}\right), 
    \end{aligned}
\end{align}
where
\begin{align}
    \begin{aligned}
\mathcal{A} & \coloneqq\left(\begin{array}{ll}
\partial E_{Q} / \partial\left(\partial_t Q\right) & \partial E_{Q} / \partial\left(\partial_t P\right) \\
\partial E_{P} / \partial\left(\partial_t Q\right) & \partial E_{P} / \partial\left(\partial_t P\right)
\end{array}\right), \\
\mathcal{B} & \coloneqq\left(\begin{array}{ll}
\partial E_{Q} / \partial\left(\partial_r Q\right) & \partial E_{Q} / \partial\left(\partial_r P\right) \\
\partial E_{P} / \partial\left(\partial_r Q\right) & \partial E_{P} / \partial\left(\partial_r P\right)
\end{array}\right), \\
\mathcal{Q} & \coloneqq\left(\begin{array}{ll}
\partial E_{Q} / \partial\left(\partial_r A\right) & \partial E_{Q} / \partial\left(\partial_r B\right) \\
\partial E_{P} / \partial\left(\partial_r A\right) & \partial E_{P} / \partial\left(\partial_r B\right)
\end{array}\right), \\
\mathcal{R} & \coloneqq\left(\begin{array}{ll}
\partial C_{A} / \partial\left(\partial_r Q\right) & \partial C_{A} / \partial\left(\partial_r P\right) \\
\partial C_{\zeta} / \partial\left(\partial_r Q\right) & \partial C_{\zeta} / \partial\left(\partial_r P\right)
\end{array}\right), \\
\mathcal{S} & \coloneqq\left(\begin{array}{ll}
\partial C_{A} / \partial\left(\partial_r A\right) & \partial C_{A} / \partial\left(\partial_r \zeta\right) \\
\partial C_{\zeta} / \partial\left(\partial_r A\right) & \partial C_{\zeta} / \partial\left(\partial_r \zeta\right)
\end{array}\right) .
\end{aligned}
\end{align}
If $\xi_{\mu}$ satisfies the characteristic equation then the corresponding values of the characteristic speeds, which are defined in spherical symmetry as $c \coloneqq -\frac{\xi_{t}}{\xi_{r}}$, are given by
\begin{align}
    \label{eq:char_speeds}
    c_{\pm} = \frac{1}{2} \left(\text{Tr}(\mathcal{C}) \pm \sqrt{\mathcal{D}}\right),
\end{align}
where, 
\begin{align}
    \mathcal{D} & \coloneqq \text{Tr}(\mathcal{C})^2 - 4\text{Det}(\mathcal{C}), \\ 
    \mathcal{C} & \coloneqq \mathcal{A}^{-1} \cdot\left(\mathcal{B}-\mathcal{Q} \cdot \mathcal{S}^{-1} \cdot \mathcal{R}\right) .
\end{align}

\section{Convergence Tests}
\begin{figure}[h!]
     \centering
    \includegraphics[width=1\linewidth]{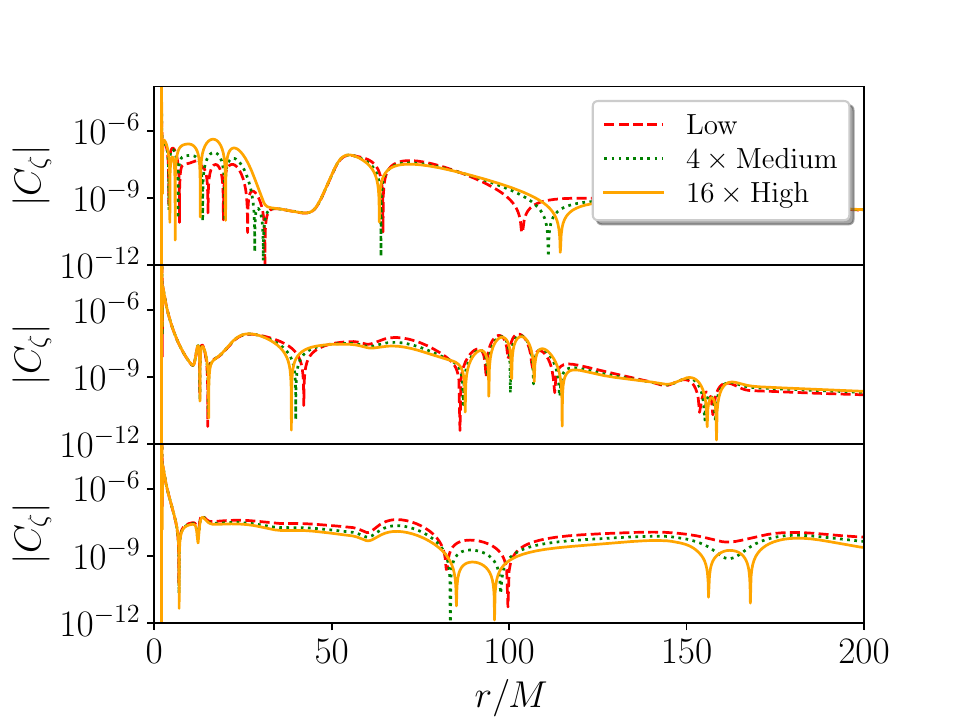}
     \caption{The convergence behaviour of the $\zeta$ constraint $C_{\zeta}$ in space for three resolutions $\Delta x=\{0.025,0.0125,0.00625\}$. Each panel represents an instance in time, the top panel is at $t/M=30$, the middle panel is at $t/M=156$, and the bottom panel is at $t/M=310$. The initial data is $a_{0} = 10^{-3}$ with $\alpha/M^2 = 1, \beta=2$. We observe second-order convergence.}
     \label{fig:convergence_test}
    \end{figure}

In this section, we present convergence tests to validate the simulations we have performed. We use three different resolutions $\{\Delta x_{\text{Low}}, \Delta x_{\text{Medium}}, \Delta x_{\text{High}}\} = \{0.025,0.0125,0.00625\}$. In Fig.~\ref{fig:convergence_test} we display second-order convergence of the $\zeta$ constraint $C_{\zeta}$ for different times during the evolution.    

\section{Linear perturbations around Schwarzschild}\label{app:linear_perturbation}
We want to study small perturbations of equation~\eqref{eq:scalar} around a Schwarzschild background to determine the timescale of the tachyonic instability. We assume a scalar field perturbation of the form
\begin{equation}
    \phi = \frac{\delta\phi(r)}{r} Y_{\ell m}(\theta, \varphi) \mathrm{e}^{\mathrm{i}\omega t} \,,
\end{equation}
for which the scalar field equation decouples from the Einstein equations and reads
\begin{equation}~\label{eq:scalarperturbation}
    \Box \, \delta \phi + \alpha f'(0) \mathcal{G}_\mathrm{Sch} \delta \phi = 0 \,,
\end{equation}
where the Gauss-Bonnet invariant for the Schwharzschild metric reads $\mathcal{G}_\mathrm{Sch} = 12/r^6$ (we use units where $M=1/2$).
We use the continued fraction method to compute the growth rates for the scalarized BHs. Let us assume the following ansatz~\cite{Leaver:1985ax}
\begin{equation}\label{ansatz}
     \delta \phi = A^{-\mathrm{i} \omega} r^{\mathrm{i} \omega -1 } \mathrm{e}^{\mathrm{i} \omega r} \sum_{n = 0}^N a_n A^n\,,
\end{equation}
where $A = g_{tt} = g^{rr} =1-1/r$.
With these definitions, equation~\eqref{eq:scalarperturbation}, after some manipulation, takes the form
\begin{equation}
    \sum_{j = -1}^{4} \left( \widetilde{\gamma}_{n,j-1} + \widetilde{\beta}_{n,j} + \widetilde{\alpha}_{n,j+1} \right) a_{n-j} = 0\,.
    \label{recurrence_relation}
\end{equation}
The coefficients appearing in the relation are given by
\begin{align}
    \widetilde{\alpha}_{n,j} & = \binom{0}{j} \alpha^r_{n-j} \,, \\
    \widetilde{\beta}_{n,j} & = \binom{0}{j} \beta^r_{n-j} + 3\alpha(-1)^j \binom{4}{j} \,, \\
    \widetilde{\gamma}_{n,j} & = \binom{0}{j} \gamma^r_{n-j} \, .
\end{align}
We notice that, from the definition of the binomial, $\widetilde{\alpha}_{n,j}$ and $\widetilde{\gamma}_{n,j}$ are non-vanishing only for $j = 0$, while $\widetilde{\beta}_{n,j}$ is non-zero for $0 \leq j \leq 4$.
The coefficients appearing in the relation are
\begin{align}
    \alpha^r_n = & \, (n+1) \left( n + 1 -2\mathrm{i} \omega \right), \\
    \beta^r_n = & \, \ell(\ell+1) + \left(2\omega + \mathrm{i} n \right)^2 - \left( n + 1 -2\mathrm{i}\omega \right)^2,   \\
    \gamma^r_n = & \, \left(n -2 \mathrm{i}\omega \right)^2 \,.
\end{align}
Notice that~\eqref{recurrence_relation} is a five-term relation, but it can be easily brought to a three-term relation through Gaussian elimination (we follow the procedure explained in the appendix of~\cite{Volkel:2022aca}). Then, it takes  the generic form
\begin{align}\label{eq:coeff_0relation}
    \beta_0 a_0 + \alpha_0 a_1 & = 0\,, \\
    \gamma_{n} a_{n-1} + \beta_{n} a_{n} + \alpha_{n} R _{n+1} & = 0 \quad \text{for } n \geq 1\,.   \label{eq:coeff_relation}
\end{align}
To invert the relation we can define the ladder operators which have the following property $a_{n+1} = -\Lambda_n a_n$, with 
\begin{equation}\label{eq:cf_relation}
    \Lambda_n = \frac{\gamma_{n+1}}{\beta_{n+1} - \alpha_{n+1} \Lambda_{n+1}}\,,
\end{equation}
where the coefficients $\alpha_n$, $\beta_n$ and $\gamma_n$ are determined by the Gaussian elimination and are not explicitly reported here. By initializing $\Lambda_N$ for a very large $N$ with the Nollert procedure~\cite{Nollert:1993zz} the equation one needs to solve to obtain the eigenfrequency $\omega$ is
\begin{align}
    \mathcal{L} = \Lambda_1 \alpha_0 - \beta_0 & = 0 \,. \label{eq:Leaver_condition_r}
\end{align}

\begin{table}[h!]
    \centering
    \begin{tabular}{cc|cc|cc}
        \multicolumn{2}{c|}{Coupling constants} & \multirow{2}*{$\omega_{I}$} & \multirow{2}*{$\omega_{I}^\mathrm{cf}$} & \multirow{2}*{$\omega_{R}$} & \multirow{2}*{$\omega_{R}^\mathrm{cf}$} \\
         $\alpha/M^2$ & $\beta$ & & & & \\
       \hline
         0.0 & 0.0 & 0.105304 & 0.104896 & 0.106749 & 0.110455 \\
         0.025 & 0.0 & 0.105012 & 0.104279 & 0.103731 & 0.106891 \\
         0.025 & 0.5 & 0.104984 & & 0.103715 & \\
         0.025 & 0.75 & 0.104968 & & 0.103703 & \\
         0.025 & 1.0 & 0.104949 & & 0.103688 & \\
         0.025 & 1.25 & 0.104929 & & 0.10367 & \\
         0.025 & 1.5 & 0.104906 & & 0.10365 & \\
         0.025 & 1.75 & 0.104881 & & 0.103626 & \\
         0.025 & 2.0 & 0.104855 & & 0.103598 & \\
         0.025 & 2.25 & 0.104829 & & 0.103565 & \\
         0.025 & 2.5 & 0.104799 & & 0.103528 & \\
         0.05 & 0.0 & 0.104481 & 0.103652 & 0.100866 & 0.103219 \\
         0.0625 & 0.0 & 0.103307 & 0.103334 & 0.10036 & 0.101338 \\
         0.075 & 0.0 & 0.103419 & 0.103011 & 0.0981844 & 0.099425 \\
         0.075 & 0.5 & 0.103394 & & 0.0981684 & \\
         0.075 & 0.75 & 0.103383 & & 0.0981544 & \\
         0.075 & 1.0 & 0.103372 & & 0.0981365 & \\
         0.075 & 1.25 & 0.103362 & & 0.0981146 & \\
         0.075 & 1.5 & 0.10335 & & 0.0980896 & \\
         0.075 & 1.75 & 0.103338 & & 0.0980609 & \\
         0.075 & 2.0 & 0.103327 & & 0.0980267 & \\
         0.075 & 2.25 & 0.10332 & & 0.0979849 & \\
         0.075 & 2.5 & 0.103317 & & 0.097936 & \\
         0.0875 & 0.0 & 0.103179 & 0.102683 & 0.096312 & 0.097478 \\
         0.1 & 0.0 & 0.10246 & 0.102350 & 0.0947566 & 0.095494 \\
         0.1125 & 0.0 & 0.102088 & 0.102011 & 0.0927956 & 0.093471 \\
         0.125 & 0.0 & 0.10167 & 0.101664 & 0.0907635 & 0.091407 \\
         0.125 & 0.5 & 0.101654 & & 0.0907499 & \\
         0.125 & 0.75 & 0.101651 & & 0.0907381 & \\
         0.125 & 1.0 & 0.101626 & & 0.0907378 & \\
         0.125 & 1.25 & 0.101615 & & 0.0907279 & \\
         0.125 & 1.5 & 0.101612 & & 0.0907116 & \\
         0.125 & 1.75 & 0.101606 & & 0.0906933 & \\
         0.125 & 2.0 & 0.101603 & & 0.0906713 & \\
         0.125 & 2.25 & 0.101602 & & 0.0906524 & \\
         0.125 & 2.5 & 0.101592 & & 0.0906586 & \\
    \end{tabular}
    \caption{The quasi-normal modes for different values of the couplings $\alpha/M^2$, and $\beta$.}
    \label{tab:QNMs}
\end{table}

\bibliography{bibnote.bib}
\end{document}